\documentclass{osa-article}
\usepackage{newunicodechar}
\newunicodechar{ }{\,}

\journal{osac}

\articletype{Research Article}
\usepackage{amsmath}
\usepackage{lineno}
\usepackage{adjustbox}
\usepackage{bm}
\usepackage{multirow}
\usepackage{wrapfig}

\begin{document}
\title{Sideband-Resolved 4H-SiC Optomechanical Resonators with Interference-Engineered Anchor-Loss Suppression}

\author{Xirui Gou \authormark{1}, William Privratsky \authormark{1}, Wenhan Sun \authormark{1}, Yuncong Liu \authormark{2}, Hamed Abiri \authormark{3}, Philip X.-L. Feng \authormark{2}, and Qing Li\authormark{1*}}

\address{\authormark{1}Electrical and Computer Engineering, Carnegie Mellon University, Pittsburgh, PA 15213, USA\\
\authormark{2}Electrical and Computer Engineering, University of Florida, Gainesville, FL 32611, USA\\
\authormark{3}School of Electrical and Computer Engineering, Georgia Institute of Technology, Atlanta, GA 30332 USA}

\email{\authormark{*}qingli2@andrew.cmu.edu} 



\begin{abstract*}
Sideband-resolved cavity optomechanical resonators provide a powerful platform for coherent photon--phonon interactions, enabling applications ranging from quantum state transduction and optomechanically induced transparency to precision sensing and microwave photonics. Achieving this regime in integrated microresonators, however, requires simultaneously realizing a narrow optical cavity linewidth, a high-frequency mechanical mode, and low mechanical dissipation. Here, we report the first sideband-resolved optomechanical resonators based on the 4H silicon carbide (4H-SiC) platform. By combining compact microdisk geometries with interference-engineered anchor-loss suppression, we simultaneously achieve intrinsic optical quality factors exceeding $1\times10^6$, room-temperature mechanical quality factors up to $1.51\times10^4$, and a sideband-resolution factor greater than seven. Systematic numerical and experimental studies reveal that a local minimum in anchor loss enables high mechanical quality factors without requiring aggressive undercutting, substantially improving fabrication yield and device robustness. We further demonstrate the first observation of optomechanically induced transparency in integrated 4H-SiC resonators, confirming coherent cavity optomechanical interactions in this material platform. These results establish 4H-SiC as a promising platform for integrated cavity optomechanics and provide a practical route toward scalable photon--phonon devices for classical and quantum photonic technologies.
\end{abstract*}

\section{Introduction}
Cavity optomechanical resonators exploit the interaction between confined optical and mechanical modes to enable coherent photon--phonon coupling through radiation pressure and photoelastic effects \cite{aspelmeyerRMP2014cavity}. Their ability to manipulate mechanical motion using light has stimulated extensive research over the past two decades, leading to numerous applications in precision sensing \cite{Painter_accelerometer, YangLan_Nature_opto_soliton}, microwave photonics \cite{Wong_phononcomb_2014, Li_Opto_Nano}, nonlinear optics \cite{Painter_OMC_microwave}, and quantum information science \cite{Painter_lasercooling_Nature, Groblacher_Micro_Optical, Loncar_phonon_NV}. A particularly important operating regime is the sideband-resolved regime, where the mechanical resonance frequency exceeds the optical cavity linewidth \cite{Kippenberg_Silica_sideband}. Under this condition, the optical cavity spectrally resolves the Stokes and anti-Stokes sidebands, enabling coherent optomechanical phenomena such as ground-state cooling \cite{Painter_lasercooling_Nature}, optomechanically induced transparency \cite{Painter_EIT, Davanco_SiN_OMC}, coherent wavelength conversion \cite{Painter_OMC_microwave}, and quantum transduction \cite{Painter_OMC_Transducer, Groblacher_Micro_Optical}.

To realize sideband-resolved operation, both a high mechanical resonance frequency and a narrow optical cavity linewidth are required. Significant progress has been achieved in recent years using a variety of integrated photonic platforms, particularly silicon optomechanical crystals, which combine strong optical and mechanical confinement to achieve exceptionally large optomechanical interactions \cite{Painter_OMC, Painter_lasercooling_Nature, Painter_EIT, Optica_clamped_OMC}. Sideband-resolved cavity optomechanics has also been demonstrated in silica \cite{Kippenberg_Silica_sideband}, silicon nitride \cite{Davanco_SiN_OMC}, diamond \cite{Barclay_diamond_disk, Loncar_diammond_OMC}, lithium niobate \cite{AmirS_LN_OMC}, gallium phosphide \cite{Schliesser_GaP_OMC}, and several other integrated photonic platforms.

Among emerging integrated photonic materials, 4H silicon carbide (4H-SiC) has attracted increasing interest because of its wide bandgap, excellent optical and mechanical properties, high thermal conductivity, and compatibility with optically addressable color centers \cite{Vuckovic_SiC_review}. These attributes have enabled rapid progress in nonlinear photonics, quantum photonics, and cavity optomechanics \cite{Awschalom_SiC_qubit, Vuckovic_4HSiC_nphoton, Li_4HSiC_comb}. Recently, high-performance suspended 4H-SiC microdisk optomechanical resonators with optical quality factors approaching one million and room-temperature mechanical quality factors exceeding $10^4$ have been demonstrated\cite{Li_4HSiC_4umDisk}. However, sideband-resolved operation has not yet been realized in this platform, primarily because increasing the mechanical resonance frequency by reducing the microdisk size often comes at the expense of increased optical radiation loss and mechanical anchor loss.

In this work, we report the first sideband-resolved optomechanical resonators based on the 4H-SiC-on-insulator (4H-SiCOI) platform. By combining compact microdisk geometries with interference-engineered anchor-loss suppression, we simultaneously achieve intrinsic optical quality factors exceeding $1\times10^6$, room-temperature mechanical quality factors up to $1.51\times10^4$, and a sideband-resolution factor exceeding seven. The resulting mechanical frequency--quality factor product is approximately $24$~THz, among the highest reported for microdisk optomechanical resonators across various integrated photonic material platforms. We further demonstrate optomechanically induced transparency, representing the first experimental observation of coherent cavity optomechanical interactions in integrated 4H-SiC microdisk resonators. The proposed anchor-loss engineering strategy enables high mechanical quality factors without requiring extreme undercut ratios, substantially improving fabrication yield while preserving excellent device performance.
\section{4H-SiC optomechanical resonator: design and demonstration}
\subsection{Device design}
\begin{figure}[ht]
  \includegraphics[width=0.96\linewidth]{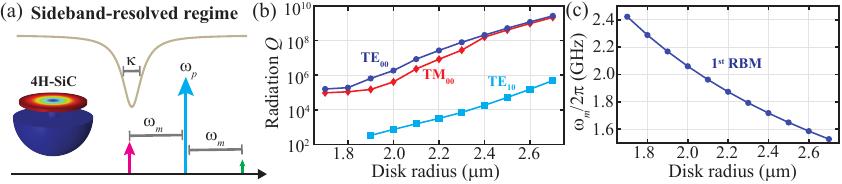}
  \caption{Sideband-resolved 4H-SiC microdisk optomechanical resonator design. (a) Schematic of a 4H-SiC microdisk resonator operated in the sideband-resolved regime, defined by $\omega_m > \kappa$, where $\omega_m$ is the mechanical resonance frequency and $\kappa$ is the optical cavity linewidth. (b) Simulated radiation-limited optical $Q$ factor in the 1550 nm band versus microdisk radius for the fundamental transverse-electric (TE$_{00}$), second-order TE$_{10}$, and fundamental transverse-magnetic (TM$_{00}$) optical modes, assuming a device layer (4H-SiC) thickness of 450 nm. 
(c) Simulated fundamental radial breathing mode (RBM) frequency of the suspended 4H-SiC microdisk as a function of disk radius.}
  \label{Fig_design}
\end{figure}
The interaction between the optical and mechanical modes in an optomechanical resonator gives rise to Stokes and anti-Stokes optical sidebands, whose frequency offsets from the pump laser are equal to the mechanical resonance frequency, $\omega_m$. Spectral resolution of these sidebands is achieved in the sideband-resolved regime, where the mechanical frequency exceeds the optical cavity linewidth, i.e., $\omega_m>\kappa$, with $\kappa$ denoting the optical cavity decay rate ($\kappa=\omega_\mathrm{opt}/Q_\mathrm{opt}$), as illustrated in Fig.~\ref{Fig_design}(a). Operating in this regime therefore calls for simultaneously maximizing the mechanical resonance frequency while minimizing the optical cavity linewidth. In this work, we focus on the fundamental radial breathing mode (RBM), whose radial displacement profile exhibits strong spatial overlap with the whispering-gallery optical mode, resulting in a large optomechanical coupling coefficient. Since the RBM frequency scales approximately inversely with the microdisk radius, reducing the resonator size provides a straightforward route toward increasing $\omega_m$. However, shrinking the device also exacerbates optical bending radiation loss \cite{QiangLin_Sidisk_lowpower,Barclay_diamond_disk}, resulting in a tradeoff between mechanical frequency and optical quality factor.

To identify the optimal device geometry, finite-element simulations were performed to evaluate both the optical radiation loss and the mechanical resonance frequency as functions of the disk radius. Figure~\ref{Fig_design}(b) shows the simulated radiation-limited optical quality factor, $Q_\mathrm{rad}$, for the fundamental transverse-electric (TE$_{00}$), first higher-order transverse-electric (TE$_{10}$), and fundamental transverse-magnetic (TM$_{00}$) whispering-gallery modes. The TE$_{00}$ and TM$_{00}$ modes maintain $Q_\mathrm{rad}$ exceeding $10^6$ for disk radii larger than approximately $2~\mu$m, whereas higher-order modes exhibit substantially stronger radiation loss. Figure~\ref{Fig_design}(c) shows the simulated frequency of the fundamental RBM, which increases approximately inversely with decreasing disk radius and exceeds $1.5$~GHz for radii below $2.7~\mu$m.

Based on these results, a target radius range of $2.3\ \mu$m--$2.7~\mu$m is selected for device fabrication. Within this range, the TE$_{00}$ and TM$_{00}$ modes remain on the radiation-loss-limited high-$Q$ plateau while the RBM frequency exceeds $1.5$~GHz, making sideband-resolved operation feasible. The combination of a narrow optical linewidth and a GHz-frequency mechanical mode yields a large $\omega_m/\kappa$ ratio while preserving a very small device footprint.

\subsection{Device fabrication and experimental setup}
The nanofabrication process of the 4H-SiC microdisk resonators follows a procedure similar to that reported in Ref.~\cite{Li_4HSiC_4umDisk}. The fabrication begins with a 4H-SiCOI chip consisting of a 630-nm-thick 4H-SiC device layer on top of a $2$-$\mu$m-thick silicon dioxide layer (NGK Insulators). The microdisk structures are first defined by electron-beam lithography (EBL) using a negative-tone resist (Flowable Oxide-16), followed by fluorine-based inductively coupled plasma (ICP) dry etching to remove approximately 560~nm of the SiC layer. A second EBL step employing a positive-tone resist (PMMA) is then performed to define circular release windows surrounding each microdisk. Within these windows, the surrounding SiC is etched away to expose the underlying oxide for the subsequent undercut process. Although only approximately 70~nm of SiC remains in these regions, the etch time is calibrated for approximately 160~nm of SiC to ensure its complete removal despite local etch-rate variations. This step simultaneously reduces the microdisk device thickness to approximately 470~nm. Finally, the chip is immersed in buffered oxide etchant (BOE) to isotropically remove the buried oxide beneath the microdisks. The etch duration is carefully controlled to achieve the desired undercut ratio, whose influence on the mechanical properties is discussed in later subsections.

\begin{figure}
\centering
\includegraphics[width=0.55\textwidth]{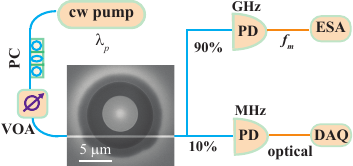}
\caption{Schematic of the experimental setup used to characterize the optical and mechanical properties of the fabricated 4H-SiC microdisk resonator. PC: polarization controller; VOA: variable optical attenuator; PD: photodetector; DAQ: data acquisition; ESA: electrical spectrum analyzer.}
\label{Fig_setup}
\end{figure}

The experimental setup used to characterize the optical and mechanical properties of the fabricated 4H-SiC microdisks is illustrated in Fig.~\ref{Fig_setup}. A continuous-wave tunable laser (Toptica CTL 1500, tuning range 1460--1570~nm) serves as the optical source, with its polarization and input power independently controlled prior to coupling into the chip. The transmitted light is divided by a $90{:}10$ fiber coupler. The $10\%$ output is directed to a low-bandwidth, high-gain photodetector (Thorlabs PDB450C) for optical transmission measurements, while the remaining $90\%$ is detected by a high-speed photodetector (Newport AD-40APD, 12-GHz bandwidth) and electrical spectrum analyzer (Tektronix RSA5106A, $6.2$-GHz bandwidth) for the mechanical mode characterization. Linear optical transmission spectra are acquired by continuously sweeping the laser wavelength across the telecom C- and L-bands while recording the photodetector signals using a data acquisition (DAQ) system.

\subsection{Optical $Q$ measurement}
\begin{figure}[ht]
  \includegraphics[width=0.96\linewidth]{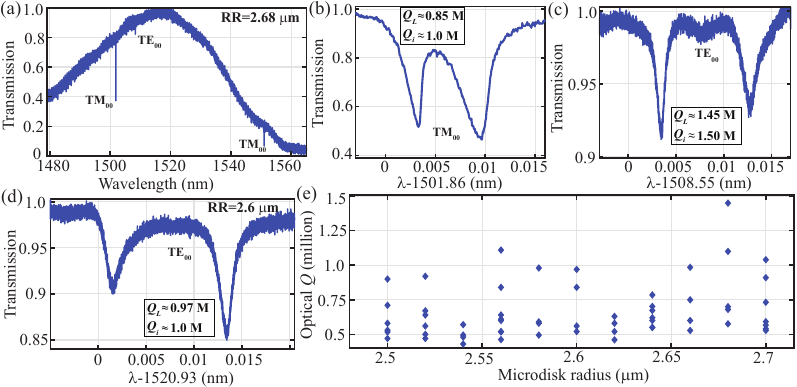}
  \caption{Optical characterization of fabricated 4H-SiC microdisk resonators.
(a) Linear swept-wavelength transmission spectrum of a $2.68$-$\mu\mathrm{m}$-radius microdisk, with zoomed-in views of the TM$_{00}$ and TE$_{00}$ resonances shown in (b) and (c), respectively.
(d) Zoomed-in TE$_{00}$ resonance of a $2.6$-$\mu\mathrm{m}$-radius microdisk.
(e) Summary of the measured loaded optical quality factors for SiC microdisks with radii ranging from $2.5~\mu\mathrm{m}$ to $2.7~\mu\mathrm{m}$.}
  \label{Fig_optical}
\end{figure}

In Fig.~\ref{Fig_optical}(a), we present a representative transmission spectrum of a 2.68-$\mu$m-radius microdisk resonator. As can be seen, only three resonance dips are observed over an 80-nm wavelength span. By comparing the measured free spectral ranges (FSRs) with finite-element simulations, these resonances are identified as the fundamental TE$_{00}$ and TM$_{00}$ whispering-gallery modes, consistent with their substantially lower radiation loss compared with higher-order resonant modes, as predicted in Fig.~\ref{Fig_design}(b). A zoomed-in view of the TM$_{00}$ resonance near $1501.86$ nm is presented in Fig.~\ref{Fig_optical}(b), yielding a loaded quality factor of \(Q_L \approx 0.85\times10^6\) and an intrinsic quality factor of \(Q_i \approx 1.00\times10^6\). Figure~\ref{Fig_optical}(c) shows the TE$_{00}$ resonance at $1508.55$ nm, exhibiting an even higher  \(Q_L \approx 1.45\times10^6\) and \(Q_i \approx 1.50\times10^6\). As another example, Fig.~\ref{Fig_optical}(d) presents the TE$_{00}$ resonance of a second microdisk with a radius of 2.6~\(\mu\)m at $1520.93$ nm, from which \(Q_L \approx 0.97\times10^6\) and \(Q_i \approx 1.00\times10^6\) are extracted.

Notably, the intrinsic quality factors extracted from measurement remain two to three orders of magnitude below the simulated radiation-limited values shown in Fig.~\ref{Fig_design}(b). This large discrepancy indicates that radiation loss is no longer the dominant limiting mechanism within the selected design window. Instead, the optical quality factor is most likely limited by fabrication-induced surface scattering and material absorption \cite{Li_4HSiC_direct_soliton}.

Figure~\ref{Fig_optical}(e) summarizes the measured loaded quality factors for an ensemble of microdisks with radii ranging from $2.5\ \mu$m to $2.7$~\(\mu\)m. Across this radius range, the devices consistently achieve loaded quality factors in the $0.5\times 10^6$ to $1\times10^6$ range, comparable to the state of the art for compact microresonators operating in the telecom band \cite{Tang_Si_disk,QiangLin_Sidisk_lowpower,Barclay_diamond_disk, Li_4HSiC_4umDisk}. These results validate the design strategy presented in Fig.~\ref{Fig_design} and demonstrate that the $2.5\ \mu$m--2.7~\(\mu\)m radius window provides a robust balance between low optical radiation loss and high mechanical resonance frequency. In contrast, microdisks with smaller radii, including those with a radius of 2.3~\(\mu\)m, exhibit lower optical quality factors (typically below \(5\times10^5\); see the Supplementary Information, SI), indicating that optical radiation loss becomes increasingly significant as the device size is further reduced.

\subsection{Anchor-loss suppression with local minimum}

\begin{figure}[ht]
  \includegraphics[width=0.96\linewidth]{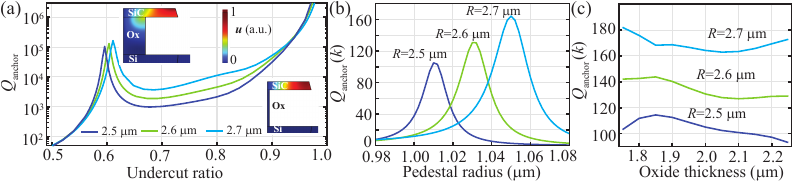}
  \caption{Simulation of anchor-loss-limited mechanical quality factor in suspended 4H-SiC microdisk resonators.
(a) Simulated anchor-loss-limited mechanical quality factor, $Q_\mathrm{anchor}$, as a function of the undercut ratio, defined as the ratio between the undercut width and the disk radius. The insets show the simulated displacement profiles of the fundamental radial breathing mode at two representative undercut ratios. (b) Zoomed-in view of the local maximum observed in (a) for a pedestal radius near $1~\mu\mathrm{m}$. (c) Simulated peak $Q_\mathrm{anchor}$ as a function of the underlying oxide-layer thickness.}
  \label{Fig_ahch_sim}
\end{figure}

In addition to achieving low optical loss, an optomechanical resonator must also minimize mechanical dissipation. For the RBM of a pedestal-supported microdisk, mechanical dissipation is often dominated by anchor loss, whereby elastic energy radiates through the supporting pedestal into the substrate \cite{Kippenberg_Silica_sideband,Tang_Si_disk}. Suppressing this loss is therefore essential for realizing high mechanical quality factors. Conventional approaches primarily rely on aggressively undercutting the microdisk to minimize the pedestal size and thereby reduce mechanical energy leakage \cite{Tang_Si_disk, Lin_3CSiC}. Although effective, this strategy is difficult to implement reproducibly, as the fragile pedestal significantly degrades the mechanical robustness of the device and often leads to structural collapse during fabrication, resulting in compromised device yield \cite{Kippenberg_Silica_sideband}.

In our recent work, however, we experimentally discovered an alternative operating regime for 4H-SiC microdisk resonators, in which the anchor loss can be strongly suppressed even with a relatively large pedestal \cite{Li_Opto_Nano}. In this regime, only approximately $50\%$--$60\%$ of the buried oxide beneath the microdisk is removed, allowing the remaining pedestal to simultaneously provide excellent mechanical support and low anchor loss. To elucidate the physical origin of this phenomenon, in this work we carried out a systematic numerical and experimental investigation of the anchor-loss behavior as a function of the undercut geometry.

Figure~\ref{Fig_ahch_sim}(a) presents the simulated anchor-loss-limited mechanical quality factor, $Q_\mathrm{anchor}$, as a function of the undercut ratio, defined as the undercut width divided by the microdisk radius. For microdisks with radii between $2.5\ \mu$m and $2.7~\mu$m, a pronounced local maximum is observed near an undercut ratio of approximately $60\%$, where $Q_\mathrm{anchor}$ increases by more than an order of magnitude compared with neighboring undercut conditions. Although similar features have been observed in previous numerical studies \cite{Kippenberg_Silica_sideband, Tang_Si_disk}, they have not been intentionally exploited to achieve low anchor loss at modest undercut ratios, nor have they been systematically investigated experimentally.  This local maximum forms the basis of our anchor-loss engineering strategy, enabling high mechanical quality factors without relying on an aggressively undercut pedestal. A magnified view of the resonance is shown in Fig.~\ref{Fig_ahch_sim}(b), corresponding to a pedestal radius of approximately $1~\mu$m. While the resonance is relatively sharp, the allowable pedestal-radius variation for maintaining $Q_\mathrm{anchor}>10^4$ exceeds $50$~nm, providing sufficient fabrication tolerance for reliable implementation.

Finally, Fig.~\ref{Fig_ahch_sim}(c) shows that the peak value of $Q_\mathrm{anchor}$ is largely insensitive to the thickness of the buried oxide layer. This observation suggests that the local maximum cannot be explained by quarter-wave interference along the vertical direction of the pedestal. Instead, we attribute the enhanced mechanical quality factor to destructive interference between laterally propagating elastic waves \cite{Tang_Si_disk}, which suppresses the net mechanical energy radiated into the substrate.

\subsection{Mechanical $Q$ measurement}

\begin{figure}[ht]
\centering
  \includegraphics[width=0.75\linewidth]{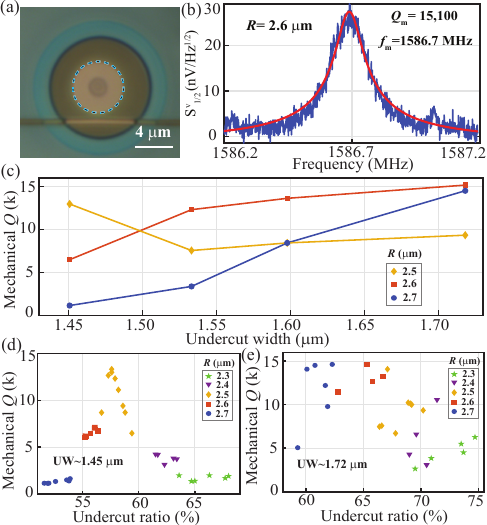}
  \caption{Experimental characterization of anchor-loss suppression in fabricated 4H-SiC microdisk optomechanical resonators.
(a) Optical micrograph of a $2.6$-$\mu\mathrm{m}$-radius 4H-SiC microdisk with an undercut ratio of approximately $60\%$.
(b) Measured electrical spectrum corresponding to the fundamental radial breathing mode of the microdisk with its optical response provided in Fig.~3(d), showing a mechanical resonance frequency of $1.586~\mathrm{GHz}$ and a room-temperature mechanical quality factor of $Q_\mathrm{m}$ =$15{,}100$ in ambient conditions.
(c) Measured mechanical quality factors of three microdisk resonators with different radii after four consecutive undercutting processes, where the undercut width increases from $1.45~\mu\mathrm{m}$ after the first undercut to $1.72~\mu\mathrm{m}$ after the fourth undercut.
(d) Summary of the measured mechanical quality factors after the first undercut for a broader set of microdisk resonators with radii ranging from $2.3~\mu\mathrm{m}$ to $2.7~\mu\mathrm{m}$.
(e) Summary of the measured mechanical quality factors for the same set of microdisk resonators after the fourth undercut. UW: Undercut width.}
  \label{Fig_mechanical}
\end{figure}

The mechanical properties of the pedestal-engineered 4H-SiC microdisks are characterized using the experimental setup shown in Fig.~\ref{Fig_setup}, where the electrical spectrum analyzer (ESA) measures the thermal Brownian motion of the mechanical mode transduced onto the transmitted optical signal. To eliminate optomechanical dynamical back-action, the optical power coupled into the cavity is kept sufficiently low, ensuring that the measured mechanical linewidth and quality factor represent the intrinsic properties of the resonator.

Figure~\ref{Fig_mechanical}(a) shows an optical micrograph of a $2.6$-$\mu$m-radius microdisk with an undercut ratio of approximately $60\%$ (whose optical property is shown in Fig.~3(d)). The corresponding ESA spectrum, shown in Fig.~\ref{Fig_mechanical}(b), exhibits a pronounced mechanical resonance at $f_m \approx 1.586$~GHz with a room-temperature mechanical quality factor of $Q_m \approx 15{,}100$ measured under ambient conditions. The measured resonance frequency agrees well with the simulated $1/R$ scaling presented in Fig.~\ref{Fig_design}(c). Importantly, the resulting $f_m\cdot Q_m\approx 23.96$~THz represents, to the best of our knowledge, the highest value reported for undercut microdisk optomechanical resonators across integrated photonic material platforms \cite{Barclay_diamond_disk, Li_4HSiC_4umDisk}.

The effectiveness of the proposed anchor-loss engineering strategy is experimentally validated in Fig.~\ref{Fig_mechanical}(c). The mechanical quality factors of three representative microdisks with different radii are monitored over four successive BOE undercut steps. In each step, the cumulative etch time is increased, expanding the undercut width from approximately $1.45~\mu$m after the first undercut to $1.72~\mu$m after the fourth. Rather than exhibiting a monotonic dependence on the undercut width, the measured $Q_m$ values show distinctly different trends for the three microdisks, reflecting their different undercut ratios. This non-monotonic behavior is in excellent qualitative agreement with the simulated $Q_\mathrm{anchor}$ shown in Figs.~\ref{Fig_ahch_sim}(a) and \ref{Fig_ahch_sim}(b). Furthermore, the observation that microdisks with different radii achieve their maximum $Q_m$ after different undercut steps is consistent with the radius-dependent optimum undercut ratio predicted by the simulations.

Figures \ref{Fig_mechanical}(d) and \ref{Fig_mechanical}(e) further summarize the measured $Q_m$ values for a larger ensemble of microdisks with radii ranging from $2.3\ \mu$m to $2.7~\mu$m after the first and fourth undercut processes, respectively. As predicted by the anchor-loss simulations, the larger-radius devices ($R\ge2.6~\mu$m) initially lie on the left side of local maximum and evolve toward the optimum undercut condition after successive BOE etching, resulting in substantially improved mechanical quality factors. In contrast, the $2.5$-$\mu$m-radius devices are already close to the local maximum after the first undercut and therefore exhibit a reduction in $Q_m$ following additional oxide removal. Smaller-radius devices gradually approach the elevated tail at much larger undercut ratios (see Fig.~4(a)), leading to a modest recovery of their mechanical quality factors.

Overall, mechanical quality factors exceeding $10^4$ are reproducibly obtained across a broad range of device radii by operating near the local $Q_\mathrm{anchor}$ maximum. Because this operating point retains a relatively large supporting pedestal, the microdisks remain mechanically robust, and the pedestal dimensions can be accurately monitored using conventional optical microscopy throughout fabrication. By comparison, achieving a similar mechanical quality factor through aggressive undercutting requires undercut ratios approaching $90\%$, leaving an extremely thin pedestal. In this regime, a large fraction of the devices collapse during wet etching or drying, reducing the fabrication yield to below $20\%$. We further note that the attainable mechanical quality factor remains comparable in this deeply undercut regime (see SI), suggesting that the experimentally observed $Q_m$ of approximately $1.5\times10^4$ is no longer limited by anchor loss. Instead, other dissipation mechanisms, such as surface-related losses, are likely to dominate the mechanical damping. These results demonstrate that interference-engineered anchor-loss suppression provides an effective strategy for simultaneously achieving high mechanical quality factors and high fabrication yield.

\section{Application: OMIT demonstration}
\begin{figure}[ht]
\centering
  \includegraphics[width=1.0\linewidth]{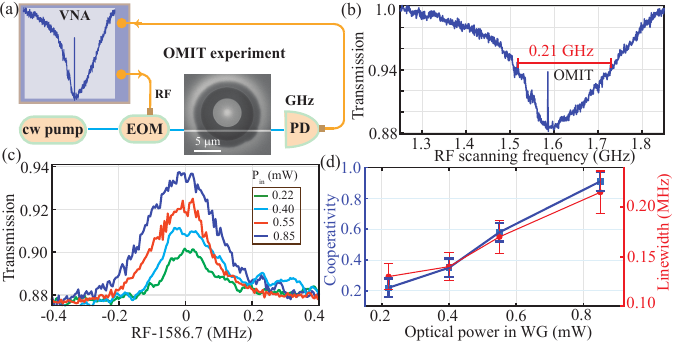}
  \caption{
Optomechanically induced transparency (OMIT) measurement.
(a) Experimental setup employing an electro-optic modulator (EOM) and a vector network analyzer (VNA).
(b) Representative VNA transmission trace of a $2.6$-$\mu$m-radius 4H-SiC microdisk (whose optical and mechanical properties are shown in Figs.~3(d) and 5(b), respectively) measured with an on-chip optical power of $0.85$ mW. The broad resonance envelope corresponds to the optical cavity resonance, while a narrow OMIT transparency peak emerges at the mechanical resonance frequency.
(c) Enlarged view of the OMIT transparency window measured under four different on-chip optical powers.
(d) Extracted optomechanical cooperativity (left axis) and measured OMIT linewidth (right axis) as functions of the on-chip optical power.
}
  \label{Fig_OMIT}
\end{figure}

The optical and mechanical characterization presented in the previous section establishes that the fabricated 4H-SiC microdisk resonators operate in the sideband-resolved regime (\(\omega_m>\kappa\)). Under this condition, the optical cavity spectrally resolves the Stokes and anti-Stokes sidebands generated by the mechanical motion, as illustrated in Fig.~\ref{Fig_design}(a). A hallmark of the sideband-resolved regime, as well as a powerful technique for characterizing cavity optomechanical interactions, is optomechanically induced transparency (OMIT), the optomechanical counterpart of electromagnetically induced transparency (EIT) in atomic systems \cite{Painter_EIT}. As already discussed in detail in the literature \cite{Kippenberg_OMIT, Painter_EIT}, OMIT originates from the coherent interference between two optical pathways mediated by a common mechanical mode.

The experimental configuration is illustrated in Fig.~\ref{Fig_OMIT}(a). A strong control laser is tuned to the red mechanical sideband of an optical resonance, i.e., detuned from the cavity resonance by one mechanical frequency (\(\Delta=\omega_m\)). An electro-optic modulator (EOM) generates weak optical sidebands around the control laser, one of which serves as a coherent probe by sweeping across the cavity resonance through continuous variation of the RF modulation frequency. In the absence of optomechanical coupling, the detected transmission simply follows the optical cavity response. When the optomechanical interaction is present, however, the probe field can also be generated through anti-Stokes scattering of the control laser by the mechanical resonator. The interference between this anti-Stokes field and the directly transmitted probe produces a narrow transparency window within the optical resonance, giving rise to the characteristic OMIT feature observed in transmission.

Figure~\ref{Fig_OMIT}(b) shows the measured VNA transmission trace for the $2.6$-$\mu$m-radius microdisk whose optical and mechanical properties were characterized in Figs.~3(d) and 5(b), respectively. The broad resonance exhibits a full width at half maximum (FWHM) of approximately $0.21$~GHz, in good agreement with the optical linewidth inferred from the measured cavity quality factor in Fig.~3(d). When the control laser is tuned to the red mechanical sideband with an estimated on-chip power of $0.85$~mW, a pronounced OMIT transparency window emerges at the center of the optical resonance. Figure~\ref{Fig_OMIT}(c) presents an enlarged view of the OMIT feature measured under four different on-chip optical powers, clearly showing that both the transparency depth and linewidth increase with increasing optical power.

The measured OMIT spectra can be fitted using the standard cavity optomechanics model to extract the optomechanical cooperativity:
\[
C=\frac{4g_0^2n_c}{\kappa\gamma_m}=C_0n_c,
\]
where $g_0$ is the vacuum optomechanical coupling rate, $n_c$ is the intracavity photon number, $\kappa$ is the optical cavity decay rate, $\gamma_m$ is the mechanical energy decay rate, and $C_0\equiv4g_0^2/\kappa\gamma_m$ is the single-photon cooperativity. The extracted cooperativity and the corresponding OMIT linewidth are summarized in Fig.~\ref{Fig_OMIT}(d), showing a reasonable agreement with the theoretical model (see SI for details). For example, at the maximum input waveguide power of $0.85$~mW (limited by the laser power and insertion loss), the cooperativity is measured to be $C\approx0.88$, corresponding to an estimated intracavity photon number of approximately $(2700\pm 200)$. This yields a single-photon cooperativity of $C_0\approx3.0\times10^{-4}$ and an equivalent vacuum optomechanical coupling rate of $g_0/2\pi\approx40$~kHz, in good agreement with both theoretical simulations and independent experimental estimates based on phonon-lasing measurements \cite{Li_Opto_Nano}. To the best of our knowledge, this work represents the first experimental demonstration of optomechanically induced transparency in an integrated 4H-SiC photonic platform. 

\section{Discussions}
\begin{table*}[ht]
\centering
\caption{
Representative sideband-resolved optomechanical resonators across different material platforms.
\\
Notes: Values in the $Q_m$ and $C_0$ columns marked with an asterisk (*) and a dagger ($^\dagger$) were measured under cryogenic vacuum and room-temperature vacuum conditions, respectively. All other values were measured at room temperature under ambient conditions.
}
\label{tab:sideband_resolved_comparison}
\begin{adjustbox}{max width=\textwidth}
\begin{tabular}{llllllll l}
\hline
Material & References & Device & $\kappa/2\pi$ (GHz) & $f_m$ (GHz) & $Q_m$ (k) & $g_0/2\pi$ (kHz) &$2\pi f_m/\kappa$ & $C_0$ \\
\hline
Si 
& Chan et al.~\cite{Painter_lasercooling_Nature} 
& OMC 
& $0.5$ 
& $3.68$ 
& $100^{*}$ 
&910 & $7.4$ & $0.19^*$\\

SiO$_2$ 
& Schliesser et al.~\cite{Kippenberg_Silica_sideband} 
& Microtoroid 
& $0.0032$ 
& $0.0735$ 
& $30^{\dagger}$  
& $0.86$ & $23$ 
& $3.8\times 10^{-4\dagger}$\\

Si$_3$N$_4$ 
& Davan\c{c}o et al.~\cite{Davanco_SiN_OMC} 
& OMC 
& $2.4$ 
& $3.8$ 
& $3$ 
& $133.6$ & $1.6$ 
& $2.3\times 10^{-5}$\\

LiNbO$_3$ 
& Jiang et al.~\cite{AmirS_LN_OMC} 
& OMC 
& $0.78$ 
& $2.1$ 
& $17^*$ 
& 120 & $2.7$ 
& $6\times 10^{-4*}$ \\

Diamond 
& Burek et al.~\cite{Loncar_diammond_OMC} 
& OMC 
& $1.1$ 
& $9.45$ 
& $7.7$
& 217 & $8.5$ 
& $1.2\times 10^{-4}$ \\

GaP 
& Tamaki et al.~\cite{Schliesser_GaP_OMC} 
& OMC 
& $2.5$ 
& $7.65$ 
& $1.56$
& 450 & $3.1$ 
& $6.6\times 10^{-5}$\\

\textbf{4H-SiC} 
& \textbf{This work} 
& \textbf{Microdisk} 
& $0.21$
& $1.587$ 
& $15.1$
& 40 &$7.6$ 
& $3.0\times 10^{-4}$\\
\hline
\end{tabular}
\end{adjustbox}
\end{table*}

To benchmark the performance of our sideband-resolved 4H-SiC microdisk resonators, Table~I compares the key metrics of representative integrated optomechanical platforms, including the optical cavity linewidth, mechanical resonance frequency and quality factor, sideband-resolution factor, and single-photon cooperativity. As shown, most previous demonstrations of sideband-resolved cavity optomechanics have relied on optomechanical crystals, where strong optical and mechanical confinement simultaneously enhances the mechanical resonance frequency and the vacuum optomechanical coupling strength. The primary exception is the pioneering silica microtoroid platform \cite{Kippenberg_Silica_sideband}, which achieves sideband-resolved operation through its ultrahigh optical quality factor and correspondingly narrow cavity linewidth.

In contrast, our work establishes pedestal-supported 4H-SiC microdisk resonators as a competitive alternative for sideband-resolved cavity optomechanics. Despite their comparatively simple geometry, the devices simultaneously achieve million-level optical quality factors, GHz-frequency mechanical modes, and room-temperature mechanical quality factors exceeding $1.5\times10^4$, resulting in an $f_mQ_m$ product of approximately $24$~THz, among the highest reported for suspended microdisk optomechanical resonators. More importantly, these performances are enabled by an interference-engineered anchor-loss suppression mechanism that eliminates the need for extreme undercut ratios. By operating near the local minimum in anchor loss, the resonators maintain a mechanically robust supporting pedestal while suppressing anchor loss, substantially improving fabrication yield without compromising device performance. Notably, despite the comparatively modest vacuum optomechanical coupling rate of the present microdisk geometry, the extracted single-photon cooperativity is already comparable to those reported for sideband-resolved optomechanical platforms. Beyond cavity optomechanics, the combination of a million-level optical quality factor and a compact optical mode volume enables an estimated Purcell enhancement factor exceeding $5000$ (see Table~S1 in the Supporting Information). Together, these attributes—including high optical quality, small optical mode volume, appreciable single-photon cooperativity, low mechanical dissipation, and scalable fabrication—establish the 4H-SiC microdisk platform as a promising building block for integrated cavity optomechanics, hybrid quantum systems, and quantum photonic technologies.

\section{Conclusion}

In summary, we have demonstrated the first sideband-resolved optomechanical resonators based on the 4H-SiC platform. By combining compact microdisk geometries with interference-engineered anchor-loss suppression, the fabricated devices simultaneously achieve intrinsic optical quality factors exceeding one million, room-temperature mechanical quality factors up to $1.51\times10^4$, and a sideband-resolution factor greater than seven. The proposed anchor-loss engineering strategy enables high mechanical quality factors without requiring aggressive undercutting, thereby substantially improving fabrication yield and device robustness. Furthermore, we experimentally demonstrate optomechanically induced transparency, confirming coherent photon--phonon interactions in integrated 4H-SiC microdisk resonators. These results establish 4H-SiC as a promising integrated platform for cavity optomechanics and provide a practical route toward scalable optomechanical devices for classical and quantum photonic technologies.


\medskip
\textbf{Acknowledgements} \par 
The CMU team was supported by NSF (2427228) and DARPA (HR0011-26-9-E122). The authors acknowledge the use of Bertucci Nanotechnology Laboratory at Carnegie Mellon University supported by grant BNL-78657879, and the Materials Characterization Facility supported by grant MCF-677785. W.~Sun also acknowledges the support of Nicholas Minnici Dean's Graduate Fellowship from CMU. Y.~Liu and P.~Feng at UF are thankful to the partial support from NSF IUCRC MIST Center (EEC-1939009), NSF Central Florida Semiconductor Innovation Engine (ITE-2315320), and DARPA OpTIm Program (HR00112320028).

\medskip

%





\newpage
\title{Supplementary Information}
\setcounter{equation}{0}
\setcounter{figure}{0}
\setcounter{table}{0}
\setcounter{section}{0}
\renewcommand{\thetable}{S\arabic{table}}
\renewcommand{\thefigure}{S\arabic{figure}}
\renewcommand{\theequation}{S\arabic{equation}}
\noindent 
\section{Optical quality factors of smaller microdisks}
As discussed in the main text, 4H-SiC microdisk resonators with radii between $2.5~\mu$m and $2.7~\mu$m consistently achieve loaded optical quality factors near or above one million, providing an excellent balance between low optical loss and high mechanical resonance frequency. To further investigate the trade-off between device size and optical performance, we fabricated and characterized an additional set of smaller microdisk resonators with radii ranging from $1.8~\mu$m to $2.3~\mu$m.

\begin{figure}[b]
\centering
  \includegraphics[width=0.96\linewidth]{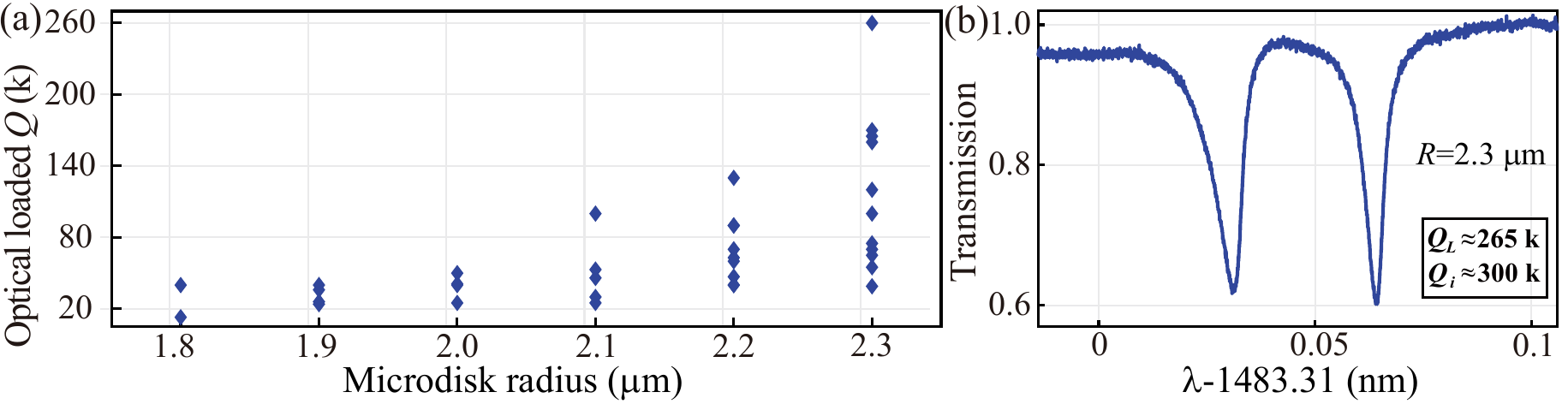}
  \caption{(a) Summary of the measured loaded optical quality factors, $Q_\mathrm{L}$, for SiC microdisks with radii ranging from $1.8~\mu\mathrm{m}$ to $2.3~\mu\mathrm{m}$. (b) Zoomed-in TE$_{00}$ resonance of a $2.3$-$\mu\mathrm{m}$-radius microdisk.}
  \label{FigS_small_optQ}
\end{figure}

Figure~\ref{FigS_small_optQ}(a) summarizes the measured loaded optical quality factors of the fundamental transverse-electric (TE$_{00}$) and transverse-magnetic (TM$_{00}$) modes for this device set. For resonators with a radius of $2.3~\mu$m, loaded quality factors exceeding $2\times10^5$ are still obtained. As an example, Fig.~\ref{FigS_small_optQ}(b) shows a TE$_{00}$ resonance of a $2.3$-$\mu$m-radius microdisk with $Q_\mathrm{L}\approx2.65\times10^5$ and $Q_\mathrm{i}\approx3.0\times10^5$. In contrast, when the radius is reduced below $2~\mu$m, the measured loaded quality factors decrease to below $4\times10^4$. This pronounced degradation is consistent with the simulated radiation-limited behavior shown in Fig.~1(b) of the main text. As the resonator size decreases, optical bending radiation loss increases rapidly. Furthermore, fabrication-induced imperfections, including sidewall roughness and surface absorption, become increasingly detrimental because the optical mode is more tightly confined near the resonator boundary, resulting in stronger interaction with surface defects.

These results confirm the design trade-off inherent in compact microdisk resonators: while reducing the disk radius increases the mechanical resonance frequency and enhances the sideband-resolution factor, it simultaneously degrades the optical quality factor through enhanced radiation loss and increased sensitivity to fabrication imperfections. Based on this characterization, we selected the $2.5$--$2.7~\mu$m radius range for the sideband-resolved devices reported in the main text, as this window optimally balances the competing requirements of high optical quality factor and GHz-frequency mechanical modes.

\section{Mechanical $Q$ with large undercut ratio}
\begin{figure}[ht]
\centering
  \includegraphics[width=0.9\linewidth]{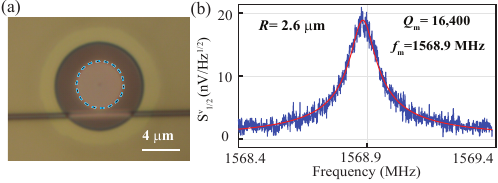}
  \caption{(a) Optical micrograph of a $2.6$-$\mu\mathrm{m}$-radius 4H-SiC microdisk after aggressive undercutting, with an estimated undercut ratio exceeding $85\%$. (b) Measured mechanical spectrum of the microdisk in (a) at room temperature under ambient conditions, showing a mechanical quality factor of approximately $16{,}400$.}
  \label{FigS_deepcut}
\end{figure}

In this work, two nominally identical 4H-SiC chips were fabricated and characterized. One chip was designed to achieve moderate undercut ratios of approximately $50$--$70\%$, and the corresponding results are presented in the main text. The second chip was subjected to more aggressive undercutting, yielding undercut ratios exceeding $80\%$. Devices on this chip exhibited a significantly lower fabrication yield, as many microdisks collapsed when the undercut ratio approached or exceeded $90\%$. Nevertheless, several surviving devices were successfully characterized. Figure~\ref{FigS_deepcut} shows one representative example, in which the supporting pedestal is almost completely removed, corresponding to an estimated undercut ratio greater than $85\%$. Despite the substantially reduced pedestal, the measured mechanical quality factor is comparable to that obtained for the moderately undercut devices operating near the local resonance peak. This observation suggests that the maximum measured mechanical quality factor of approximately $1.5\times10^4$--$1.6\times10^4$ is unlikely to be limited by anchor loss. Instead, other dissipation mechanisms, such as surface-related losses, are likely to dominate the mechanical damping.

\section{OMIT modeling}
For the OMIT measurements, the control laser is red-detuned from the optical resonance by exactly one mechanical frequency ($\Delta=\omega_m$). Under this condition, the anti-Stokes scattering process, in which a pump photon absorbs a phonon, is resonantly enhanced by the optical cavity, whereas the Stokes scattering process is strongly suppressed. This results in a beam-splitter-like optomechanical interaction, giving rise to optomechanically induced transparency (OMIT). When a weak probe field is applied, the probe transmission amplitude is given by~\cite{Painter_EIT}
\begin{equation}
t(\Omega) = 1 - \kappa_{ex}\frac{\frac{\gamma_m}{2} - i\Omega}{\left(\frac{\kappa}{2} - i\Omega\right)\left(\frac{\gamma_m}{2} - i\Omega\right) + G^2},
\end{equation}
\noindent where $\kappa_{ex}$ and $\kappa$ denote the external coupling rate and total optical decay rate, respectively, $\gamma_m=\omega_m/Q_m$ is the mechanical damping rate, $\Omega=\omega_{\mathrm{probe}}-\omega_{\mathrm{pump}}-\omega_m$ is the two-photon detuning from the OMIT resonance, and $G=g_0\sqrt{n_c}$ is the enhanced optomechanical coupling rate, with $g_0$ the vacuum optomechanical coupling rate and $n_c$ the intracavity photon number.

At exact two-photon resonance ($\Omega=0$), the transmission reduces to
\begin{equation}
T(\Omega=0)=\left|1-\frac{2\eta}{1+C}\right|^2,
\label{EqS_Tmin}
\end{equation}
\noindent where $\eta=\kappa_{ex}/\kappa$ is the optical coupling ratio and $C=C_0n_c$ is the cooperativity, with the vacuum cooperativity defined as
\begin{equation}
C_0=\frac{4g_0^2}{\kappa\gamma_m}. \label{EqS_C0}
\end{equation}
\noindent Equation~\ref{EqS_Tmin} shows that, in the absence of optomechanical coupling ($C=0$), the probe transmission reaches the minimum of the passive cavity resonance. As the cooperativity increases, destructive interference between the directly transmitted probe and the anti-Stokes-generated probe gives rise to an increasingly pronounced transparency window, allowing $C$ to be extracted directly from the transmission contrast.

Similarly, the full width at half maximum (FWHM) of the OMIT transparency window is given by~\cite{Kippenberg_OMIT}
\begin{equation}
\Gamma_{\mathrm{OMIT}}=\gamma_m(1+C),
\label{EqS_BW}
\end{equation}
\noindent which provides an independent method for determining the cooperativity from the measured linewidth broadening.

In Fig.~6(d) of the main text, the cooperativity extracted from the transmission contrast using Eq.~\ref{EqS_Tmin} (left axis) agrees well with that obtained from the OMIT linewidth using Eq.~\ref{EqS_BW} (right axis), confirming the consistency of these two independent extraction methods. Furthermore, the cooperativity can be estimated independently using Eq.~\ref{EqS_C0}, since the vacuum optomechanical coupling rate $g_0$ of these compact microdisks has previously been determined through phonon-lasing measurements \cite{Li_Opto_Nano}, while the intracavity photon number $n_c$ can be calculated from the optical power coupled into the waveguide and pump laser detuning (off resonance by the mechanical frequency). As summarized in Table~S1, the theoretically estimated cooperativity is in good agreement with the experimentally extracted values, further validating our analysis.

\section{Potential Purcell enhancement}
In addition to cavity optomechanics, Table~S1 also lists the optical mode volume of the TE$_{00}$ mode in the $2.6$-$\mu$m-radius 4H-SiC microdisk, which is estimated to be approximately $2.26~\mu\text{m}^3$. Combined with the loaded optical quality factor approaching $1\times10^6$, the resonator is expected to provide substantial Purcell enhancement for embedded quantum emitters. The Purcell factor is given by
\begin{equation}
F_p=\frac{3}{4\pi^2}\left(\frac{\lambda}{n}\right)^3\frac{Q_L}{V_{\mathrm{opt}}},
\label{EqS_Purcell}
\end{equation}
\noindent where $n$ is the refractive index of 4H-SiC ($\approx2.6$ at 1550~nm). Using the measured device parameters, Eq.~\ref{EqS_Purcell} yields a Purcell factor exceeding $5\times10^3$. Such a large Purcell enhancement, together with the demonstrated sideband-resolved optomechanical performance, makes these compact 4H-SiC microdisks a promising platform for hybrid quantum photonic systems integrating photons, phonons, and solid-state quantum emitters.

\begin{table}[ht]
\caption{Summary of the key parameters of the $2.6$-$\mu$m-radius microdisk used for the OMIT measurements presented in Fig.~6 of the main text. The control laser is red-detuned from the optical resonance by one mechanical frequency ($1.587$~GHz), and this detuning is taken into account when estimating the intracavity photon number.}
\begin{adjustbox}{width=1.0\columnwidth,center}
\begin{tabular}{c|c|c|c}
\hline
\textbf{Domain}& \textbf{Parameters}              & \textbf{Description}                & \textbf{Values}   \\ \hline
\multirow{6}{*}{Optical}&$\omega_{opt} / 2\pi$                & Optical frequency                   & $197.25$ THz \\
& $Q_L$ | $Q_i$ | $Q_c$& Loaded|Intrinsic|Coupling $Q$ factors & $0.97$M | $1.0$M| $31$M \\
&$\kappa$ & Total optical decay rate & $1.28\ \text{ns}^{-1}$ \\
&$\kappa_{ex}$ & External coupling decay rate & $0.04\ \text{ns}^{-1}$ \\
&$V_{opt}$ & Optical mode volume & $\approx 2.26\ \mu\text{m}^3$ \\
&$F_p$ & Purcell enhancement factor & $\approx 6500 $ \\
\hline 
\multirow{3}{*}{Mechanical}& $\Omega_m / 2\pi$& Mechanical frequency & $1.587$ GHz        \\
&$Q_m$ & Mechanical quality factor & 15,100 \\
& $\gamma_m$ & Mechanical energy decay rate& $0.66$ $\mu\text{s}^{-1}$ \\
\hline
\multirow{4}{*}{Optomechanical}& $g_0/2\pi$ & Vacuum OM coupling rate & $\approx 40$ kHz \cite{Li_Opto_Nano} \\
& $C_0$ & Single-photon cooperativity (Eq.~\ref{EqS_C0}) & $\approx 3.0\times 10^{-4}$ \\
&$n_c$ &Intracavity photon with $P_d=0.85$ mW & $\approx 2680$\\
&$C_\text{theory}$& \textbf{Theoretical} cooperativity for $P_d=0.85$ mW & $\approx 0.88$\\
\hline
\multirow{5}{*}{OMIT data} & $\eta$ & Optical coupling ratio & $\approx 0.031$\\
& $T(\Omega=0)$ & OMIT peak at $P_d=0.85$ mW &$\approx 0.935$ \\
& $\Gamma_\text{OMIT}$&OMIT FWHM at $P_d=0.85$ mW & $\approx 200$ kHz\\
&$C_\text{est1}$&\textbf{Estimated} $C$ using Eq.~{S2} from $T(\Omega=0)$& $\approx 0.876$ \\
&$C_\text{est2}$&\textbf{Estimated} $C$ using Eq.~{S4} from $\Gamma_\text{OMIT}$& $\approx 0.905$ \\
\hline
\end{tabular}
\end{adjustbox}
\end{table}

\bibliography{SiC_Ref}

\begin{thebibliography}{10}
\newcommand{\enquote}[1]{``#1''}

\bibitem{aspelmeyerRMP2014cavity}
M.~Aspelmeyer, T.~J. Kippenberg, and F.~Marquardt, \enquote{Cavity optomechanics,} {\protect\JournalTitle{Reviews of Modern Physics}} \textbf{86}, 1391--1452 (2014).

\bibitem{Painter_accelerometer}
A.~G. Krause, M.~Winger, T.~D. Blasius, Q.~Lin, and O.~Painter, \enquote{A high-resolution microchip optomechanical accelerometer,} {\protect\JournalTitle{Nature Photonics}} \textbf{6}, 768--772 (2012).

\bibitem{YangLan_Nature_opto_soliton}
J.~Zhang, B.~Peng, S.~Kim, F.~Monifi, X.~Jiang, Y.~Li, P.~Yu, L.~Liu, Y.-x. Liu, A.~Alù, and L.~Yang, \enquote{Optomechanical dissipative solitons,} {\protect\JournalTitle{Nature}} \textbf{600}, 75--80 (2021).

\bibitem{Wong_phononcomb_2014}
X.~Luan, Y.~Huang, Y.~Li, J.~F. McMillan, J.~Zheng, S.-W. Huang, P.-C. Hsieh, T.~Gu, D.~Wang, A.~Hati, D.~A. Howe, G.~Wen, M.~Yu, G.~Lo, D.-L. Kwong, and C.~W. Wong, \enquote{An integrated low phase noise radiation-pressure-driven optomechanical oscillator chipset,} {\protect\JournalTitle{Scientific Reports}} \textbf{4}, 6842 (2014).

\bibitem{Li_Opto_Nano}
X.~Gou, W.~Privratsky, W.~Sun, Y.~Liu, H.~Abiri, and Q.~Li, \enquote{Chip-{Scale} {Optomechanical} {Frequency} {Comb} with a 1–70 {GHz} {Span},} {\protect\JournalTitle{Nano Letters}} \textbf{25}, 17644--17649 (2025).

\bibitem{Painter_OMC_microwave}
K.~Fang, M.~H. Matheny, X.~Luan, and O.~Painter, \enquote{Optical transduction and routing of microwave phonons in cavity-optomechanical circuits,} {\protect\JournalTitle{Nature Photonics}} \textbf{10}, 489--496 (2016).

\bibitem{Painter_lasercooling_Nature}
J.~Chan, T.~P.~M. Alegre, A.~H. Safavi-Naeini, J.~T. Hill, A.~Krause, S.~Gröblacher, M.~Aspelmeyer, and O.~Painter, \enquote{Laser cooling of a nanomechanical oscillator into its quantum ground state,} {\protect\JournalTitle{Nature}} \textbf{478}, 89--92 (2011).

\bibitem{Groblacher_Micro_Optical}
M.~Forsch, R.~Stockill, A.~Wallucks, I.~Marinković, C.~Gärtner, R.~A. Norte, F.~van Otten, A.~Fiore, K.~Srinivasan, and S.~Gröblacher, \enquote{Microwave-to-optics conversion using a mechanical oscillator in its quantum ground state,} {\protect\JournalTitle{Nature Physics}} \textbf{16}, 69--74 (2020).

\bibitem{Loncar_phonon_NV}
G.~Joe, M.~Haas, K.~Kuruma, C.~Jin, D.~D. Kang, S.~W. Ding, C.~Chia, H.~Warner, B.~Pingault, B.~Machielse, S.~Meesala, and M.~Lončar, \enquote{Purcell-enhanced spin–phonon coupling with a single colour centre,} {\protect\JournalTitle{Nature}} \textbf{653}, 378--383 (2026).

\bibitem{Kippenberg_Silica_sideband}
A.~Schliesser, R.~Rivière, G.~Anetsberger, O.~Arcizet, and T.~J. Kippenberg, \enquote{Resolved-sideband cooling of a micromechanical oscillator,} {\protect\JournalTitle{Nature Physics}} \textbf{4}, 415--419 (2008).

\bibitem{Painter_EIT}
A.~H. Safavi-Naeini, T.~P.~M. Alegre, J.~Chan, M.~Eichenfield, M.~Winger, Q.~Lin, J.~T. Hill, D.~E. Chang, and O.~Painter, \enquote{Electromagnetically induced transparency and slow light with optomechanics,} {\protect\JournalTitle{Nature}} \textbf{472}, 69--73 (2011).

\bibitem{Davanco_SiN_OMC}
M.~Davanço, S.~Ates, Y.~Liu, and K.~Srinivasan, \enquote{{Si3N4} optomechanical crystals in the resolved-sideband regime,} {\protect\JournalTitle{Applied Physics Letters}} \textbf{104}, 041101 (2014).

\bibitem{Painter_OMC_Transducer}
S.~Sonar, U.~Hatipoglu, S.~Meesala, D.~P. Lake, H.~Ren, and O.~Painter, \enquote{High-efficiency low-noise optomechanical crystal photon-phonon transducers,} {\protect\JournalTitle{Optica}} \textbf{12}, 99--104 (2025).

\bibitem{Painter_OMC}
M.~Eichenfield, J.~Chan, R.~M. Camacho, K.~J. Vahala, and O.~Painter, \enquote{Optomechanical crystals,} {\protect\JournalTitle{Nature}} \textbf{462}, 78--82 (2009).

\bibitem{Optica_clamped_OMC}
J.~Kolvik, P.~Burger, J.~Frey, and R.~V. Laer, \enquote{Clamped and sideband-resolved silicon optomechanical crystals,} {\protect\JournalTitle{Optica}} \textbf{10}, 913--916 (2023).

\bibitem{Barclay_diamond_disk}
M.~Mitchell, B.~Khanaliloo, D.~P. Lake, T.~Masuda, J.~P. Hadden, and P.~E. Barclay, \enquote{Single-crystal diamond low-dissipation cavity optomechanics,} {\protect\JournalTitle{Optica}} \textbf{3}, 963--970 (2016).

\bibitem{Loncar_diammond_OMC}
M.~J. Burek, J.~D. Cohen, S.~M. Meenehan, N.~El-Sawah, C.~Chia, T.~Ruelle, S.~Meesala, J.~Rochman, H.~A. Atikian, M.~Markham, D.~J. Twitchen, M.~D. Lukin, O.~Painter, and M.~Lončar, \enquote{Diamond optomechanical crystals,} {\protect\JournalTitle{Optica}} \textbf{3}, 1404--1411 (2016).

\bibitem{AmirS_LN_OMC}
W.~Jiang, R.~N. Patel, F.~M. Mayor, T.~P. McKenna, P.~Arrangoiz-Arriola, C.~J. Sarabalis, J.~D. Witmer, R.~V. Laer, and A.~H. Safavi-Naeini, \enquote{Lithium niobate piezo-optomechanical crystals,} {\protect\JournalTitle{Optica}} \textbf{6}, 845--853 (2019).

\bibitem{Schliesser_GaP_OMC}
S.~Tamaki, M.~B. Kristensen, T.~Martel, R.~Braive, and A.~Schliesser, \enquote{Two-dimensional gallium phosphide optomechanical crystal in the resolved-sideband regime,} {\protect\JournalTitle{Optics Express}} \textbf{32}, 48500--48508 (2024).

\bibitem{Vuckovic_SiC_review}
D.~M. Lukin, M.~A. Guidry, and J.~Vu{\v c}kovi{\'c}, \enquote{Integrated {{quantum photonics}} with {{silicon carbide}}: challenges and {{prospects}},} {\protect\JournalTitle{PRX Quantum}} \textbf{1}, 020102 (2020).

\bibitem{Awschalom_SiC_qubit}
C.~P. Anderson, A.~Bourassa, K.~C. Miao, G.~Wolfowicz, P.~J. Mintun, A.~L. Crook, H.~Abe, J.~U. Hassan, N.~T. Son, T.~Ohshima, and D.~D. Awschalom, \enquote{Electrical and optical control of single spins integrated in scalable semiconductor devices,} {\protect\JournalTitle{Science}} \textbf{366}, 1225--1230 (2019).

\bibitem{Vuckovic_4HSiC_nphoton}
D.~M. Lukin, C.~Dory, M.~A. Guidry, K.~Y. Yang, S.~D. Mishra, R.~Trivedi, M.~Radulaski, S.~Sun, D.~Vercruysse, G.~H. Ahn, and J.~Vu{\v c}kovi{\'c}, \enquote{{{4H}}-silicon-carbide-on-insulator for integrated quantum and nonlinear photonics,} {\protect\JournalTitle{Nature Photonics}} \textbf{14}, 330--334 (2020).

\bibitem{Li_4HSiC_comb}
L.~Cai, J.~Li, R.~Wang, and Q.~Li, \enquote{Octave-spanning microcomb generation in {4H}-silicon-carbide-on-insulator photonics platform,} {\protect\JournalTitle{Photonics Research}} \textbf{10}, 870--876 (2022).

\bibitem{Li_4HSiC_4umDisk}
Y.~Liu, W.~Sun, H.~Abiri, P.~X.-L. Feng, and Q.~Li, \enquote{Ultracompact {4H}-silicon carbide optomechanical resonator with \textit{f}$_{\textrm{\textit{m}}}$ · \textit{{Q}}$_{\textrm{\textit{m}}}$ exceeding 10$^{\textrm{13}}$  {Hz},} {\protect\JournalTitle{Photonics Research}} \textbf{13}, 2531--2538 (2025).

\bibitem{QiangLin_Sidisk_lowpower}
W.~C. Jiang, X.~Lu, J.~Zhang, and Q.~Lin, \enquote{High-frequency silicon optomechanical oscillator with an ultralow threshold,} {\protect\JournalTitle{Optics Express}} \textbf{20}, 15991--15996 (2012).

\bibitem{Li_4HSiC_direct_soliton}
W.~Sun, J.~Li, R.~Wang, and Q.~Li, \enquote{Directly accessing the single-soliton state of a {Kerr} microcomb and its universal scaling law [{Invited}],} {\protect\JournalTitle{Optical Materials Express}} \textbf{14}, 2938--2948 (2024).

\bibitem{Tang_Si_disk}
X.~Sun, X.~Zhang, and H.~X. Tang, \enquote{High-{{Q}} silicon optomechanical microdisk resonators at gigahertz frequencies,} {\protect\JournalTitle{Applied Physics Letters}} \textbf{100}, 173116 (2012).

\bibitem{Lin_3CSiC}
X.~Lu, J.~Y. Lee, P.~X.-L. Feng, and Q.~Lin, \enquote{Silicon carbide microdisk resonator,} {\protect\JournalTitle{Optics Letters}} \textbf{38}, 1304--1306 (2013).

\bibitem{Kippenberg_OMIT}
S.~Weis, R.~Rivière, S.~Deléglise, E.~Gavartin, O.~Arcizet, A.~Schliesser, and T.~J. Kippenberg, \enquote{Optomechanically induced transparency,} {\protect\JournalTitle{Science}} \textbf{330}, 1520--1523 (2010).

\end{thebibliography}

\end{document}